\documentstyle[emulateapj]{article}

\lefthead{COLPI, GEPPERT \& PAGE}
\righthead{ AXPs period clustering and magnetic field decay in magnetars}

\newcommand{\be}{\begin{equation}} 
\newcommand{\ee}{\end{equation}}

\begin{document}%%%%%%%%%%%%%%%%%%%%%%%%%%%%%%%%%%%%%%%%%%%%%%%%%%%%%%%%%%%%%%%
%%
%%%%%%%%%%%%%%%%%%%%%%%%%%%%%%%%%%%%%%%%%%%%%%%%%%%%%%%%%%%%%%%%%%%%%%%%%%%%%%%
%%
\title{Period clustering of the Anomalous X-ray Pulsars and 
       magnetic field decay in magnetars}
\author{Monica Colpi$^1$, Ulrich Geppert$^2$ and Dany Page$^3$}
\affil{$^1$Dipartimento di Fisica, Universit\'a degli Studi di Milano Bicocca,
Piazza della Scienza 3, 20126 Milano, Italy; colpi@uni.mi.astro.it \\
$^2$ Astrophysikalisches Institut Potsdam, An der Sternwarte 16, D-14482, 
Potsdam, Germany; urme@aip.de \\
$^3$ Instituto de Astronom\'{\i}a, UNAM, 04510 M\'{e}xico D.F., M\'{e}xico;
page@astroscu.unam.mx} 

%%%%%%%%%%%%%%%%%%%%%%%%%%%%%%%%%%%%%%%%%%%%%%%%%%%%%%%%%%%%%%%%%%%%%%%%%%%%%%%
%%
\begin{abstract}%%%%%%%%%%%%%%%%%%%%%%%%%%%%%%%%%%%%%%%%%%%%%%%%%%%%%%%%%%%%%%%
%%
%%%%%%%%%%%%%%%%%%%%%%%%%%%%%%%%%%%%%%%%%%%%%%%%%%%%%%%%%%%%%%%%%%%%%%%%%%%%%%%
%%

We confront theoretical models for  the rotational, magnetic, and 
thermal evolution  of an ultra-magnetized  neutron star, or magnetar,
with  available data on  the Anomalous X-ray Pulsars (AXPs).   
We argue that, if the AXPs are interpreted as magnetars, their clustering
of spin periods between 6 and 12 seconds, 
observed at present in this class of objects,  
their period derivatives, 
their  thermal X-ray luminosities, 
and the association  of two of them with young supernova remnants, 
can only be
understood globally if 
the magnetic field in magnetars decays
significantly on a time scale of the order of 10$^4$ years.

\end{abstract}%%%%%%%%%%%%%%%%%%%%%%%%%%%%%%%%%%%%%%%%%%%%%%%%%%%%%%%%%%%%%%%%%


%%%%%%%%%%%%%%%%%%%%%%%%%%%%%%%%%%%%%%%%%%%%%%%%%%%%%%%%%%%%%%%%%%%%%%%%%%%%%%%
%%
%%%%%%%%%%%%%%%%%%%%%%%%%%%%%%%%%%%%%%%%%%%%%%%%%%%%%%%%%%%%%%%%%%%%%%%%%%%%%%%
%%
\section{INTRODUCTION}%%%%%%%%%%%%%%%%%%%%%%%%%%%%%%%%%%%%%%%%%%%%%%%%%%%%%%%%%
%%
%%%%%%%%%%%%%%%%%%%%%%%%%%%%%%%%%%%%%%%%%%%%%%%%%%%%%%%%%%%%%%%%%%%%%%%%%%%%%%%
%%
%%%%%%%%%%%%%%%%%%%%%%%%%%%%%%%%%%%%%%%%%%%%%%%%%%%%%%%%%%%%%%%%%%%%%%%%%%%%%%%
%%

The bright compact X-ray sources 
1E 2259+586, 4U 0142+61, 1E 1048.1-5937, 1E 1841-045 and 
1RXS J170849-40091 comprise
a small class of distinct objects (Mereghetti \& Stella 1995\markcite{MS95}), 
known as the ``Anomalous X-ray Pulsars'' (AXPs), whose  
spin periods $P$ are  clustered around 8 s, and are found to be spinning
down rapidly with 
period derivatives $\dot{P}$ in excess of $5 \cdot 10^{-13}$ s s$^{-1}$. 
Their X-ray luminosities $L_X$ in the range 
of $10^{34}-10^{36}$ erg s$^{-1}$ and their soft spectra make them
peculiar among the brighter and harder X-ray pulsators found in the
Galaxy.
The absence of a stellar companion and/or of a circumstellar disk at the
sensitivity limits of current optical/IR searches 
(Mereghetti, Israel \& Stella 1998\markcite{MIS98a};
Coe \& Pightling 1998\markcite{Coe98}) lead to the hypothesis
that AXPs are isolated neutron stars endowed
with an unusually  large magnetic field $B \gg 10^{13}$G 
(Vasisht \& Gotthelf 1997\markcite{VG97})
that is decaying in order to power the X-ray emission 
(Thompson \& Duncan 1996\markcite{TD96}). 
This hypothesis is strengthened by the association of both 
1E 2259+586 and 1E 1841-045 with young ($\lesssim 10^4$ yr) supernova
remnants (SNRs) suggesting  that AXPs are
isolated young neutron stars with hot cores.
There are several AXP candidates which await future determination
of their $\dot P$ for confirmation. 
Of these, the 7 s pulsar AX J1845-0258 
(Torii et al. 1998\markcite{Tetal98}; Gotthelf \& Vasisht
1998\markcite{GV98}) has recently been found to be centered on a SNR 
(Gaensler et al. 1999\markcite{Gaen99}),
while the enigmatic source
RX J0720.4-3125 (Haberl  et al. 1997\markcite{Hetal97})
has a period of 8.39 s but a lower X-ray luminosity for an AXP.

Duncan \& Thompson (1992\markcite{DT92}) had proposed that neutron stars with 
huge interior magnetic fields  $\sim 10^{15}-10^{16}$G can exist  
(Woltjer 1964\markcite{W64}),
and that in these magnetars the magnetic energy rather than the rotational
energy is the main source  powering their emission.
If it is true that a large magnetic field decays dramatically through
non linear processes such as ambipolar diffusion and/or a Hall cascade
(Goldreich \& Reisenegger 1992\markcite{GR92}), magnetic energy is 
continuously
dissipated into heat making the neutron
star incandescent and brighter than a coeval neutron star with a non
decaying field.
AXPs may  be a manifestation of this phenomenon as observed on the
time scales of 
$\sim 10^3 - 10^4$ yr (Thompson \& Duncan 1996\markcite{TD96}; 
Heyl \& Kulkarni 1998\markcite{HK98}, thereafter H\&K).

In the $P-\dot{P}$ diagram this handful of ``candidate'' magnetars 
occupy a narrow vertical strip along the $P$ axis in the range between
6 and 12 seconds.
The lower limit is usually considered to be
a natural consequence of the initial rapid spin-down 
due to the postulated ultra-strong magnetic field.
But then why has no AXP with longer period  been yet detected ?
In this letter we want to address this issue by exploring various
scenarios concerning 
the thermal and magnetic evolution of ultra-magnetized neutron stars
and consider their consequences on the rotational history of the star.

%%%%%%%%%%%%%%%%%%%%%%%%%%%%%%%%%%%%%%%%%%%%%%%%%%%%%%%%%%%%%%%%%%%%%%%%%%%%%%%
%%%
\section{The Constant Magnetic Field Scenario}
%%%%%%%%%%%%%%%%%%%%%%%%%%%%%%%%%%%%%%%%%%%%%%%%%%%%%%%%%%%%%%%%%%%%%%%%%%%%%%%
%%%

If we assume that  the magnetic field of the neutron star is constant, then  
magneto-dipolar radiation will cause rapid spin-down  to periods of 
\be
P \approx 8 \;\; (B/10^{15} {\rm G}) (t/10^{3} {\rm yr})^{1/2} \; {\rm s}.
\ee
Given the observed periods of AXPs, this implies that they must become 
suddenly
undetectable after a few thousand years.
A rapid fading of the thermal X-ray luminosity is predicted by cooling models
of ultra-strongly magnetized neutron stars due to the early dominance of the 
photon
cooling over the previous neutrino cooling.
The age at which the abrupt transition to photon cooling occurs depends 
sensitively
on the chemical composition of the envelope, and on the specific heat of the 
star's core
which is controlled mostly by neutron superfluidity (Page 1998\markcite{P98}).
Apparently, the most optimistic models indicate that this age is around, but 
definitely
above, $10^{4}$ yr. 
This scenario thus naturally leads to an upper observable period of AXPs of 
several
tens of seconds, considering field strengths above $10^{14}$ G.

%%%%%%%%%%%%%%%%%%%%%%%%%%%%%%%%%%%%%%%%%%%%%%%%%%%%%%%%%%%%%%%%%%%%%%%%%%%%%%%
%%%
\section{The Decaying Magnetic Field Scenarios}
%%%%%%%%%%%%%%%%%%%%%%%%%%%%%%%%%%%%%%%%%%%%%%%%%%%%%%%%%%%%%%%%%%%%%%%%%%%%%%%
%%%

A decaying magnetic field has two important effects on our concerns.
First, it will lead to an asymptotic saturation period and, second, it will 
maintain
the thermal X-ray luminosity high enough for detection during a longer time. 

The three main avenues of field decay, as described by Goldreich and 
Reisenegger
(1992\markcite{GR92}), are ambipolar
diffusion, in the irrotational and solenoidal modes, and the Hall cascade.
Ambipolar diffusion only acts in the neutron star core while the Hall cascade
can act in both the core and the crust. 
Results of the magneto-thermal evolution driven by ambipolar diffusion have 
been
presented by H\&K\markcite{HK98}.
We will call hereon avenue A for the irrotational
mode, and avenue B for the solenoidal mode of field decay.
Geppert et al. (1999\markcite{Getal99}) recently studied the case of a 
magnetic field
confined to the crust with its decay induced by the Hall cascade (our avenue 
C),
which resulted in a much faster decay than in the case of a field permeating 
the core.

\subsection{Field and period evolution}%%%%%%%%%%%%%%%%%%%%%%%%%%%%%%%%%%%%%%%%
%%%%

\begin{figure}
\epsscale{0.3}
\plotone{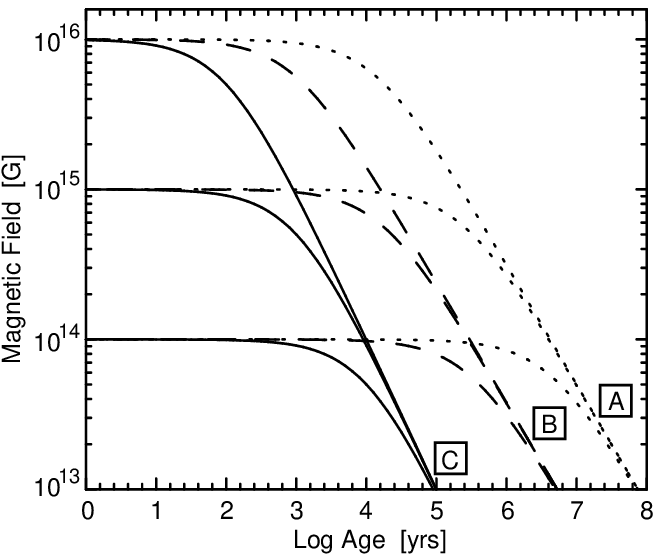}
\figcaption{ Magnetic field evolution in avenues
A (ambipolar diffusion in the irrotational mode: 
$a=0.01$ and $\alpha =5/4$),
B (ambipolar diffusion in the solenoidal mode: 
$a=0.15$ and $\alpha = 5/4$), 
and C (crustal Hall cascade: 
$a=10$ and $\alpha = 1$), 
for various initial field strengths  following equation (3).
\label{fig:1}}
\end{figure}

The results of all these numerical calculations can be well fitted by simple
decay laws of the form
\be
\frac{dB}{dt} = -a \; B^{1+\alpha}
\label{equ:decay-law}
\ee
for the surface magnetic field, which implies
\be
B(t) = \frac{B_0}{[1+a \alpha B_0^\alpha \; t]^{1/\alpha}}
\label{equ:decay}
\ee
where $B_0$ is the initial field strength that we let vary between
$10^{13}$ and $10^{16}$ G.
Figure~\ref{fig:1} shows the evolution of the magnetic field for these models
which is relevant for AXPs.
(The values of the parameters are given in the Figure 1 caption,
in units of $10^6$ yr for $t$ and $10^{13}$ G for $B$.)
The decay law (eq.[2] or [3]) holds while the current field is strong  
(i.e., $B(t)>10^{13}$ G), in  all the three avenues.
Assuming again magneto-dipolar braking, 
$P \dot{P} = b B^2$ with $b \approx 10^{-39}$ cgs
(i.e., $b \approx 3$ when $B$ is measured in 10$^{13}$ G, 
$P$ in seconds and $\dot{P}$ in seconds per 10$^6$ years),
the period evolution is then given by
\be
P^2(t) = P_0^2 + \frac{2}{(2-\alpha)} \frac{b}{a} B_0^{2-\alpha}
         \left[1-(1+a \alpha B_0^{\alpha} t)^{(\alpha -2)/\alpha}\right]
\label{equ:period}
\ee
$P_0$ being the initial period.
At old times this immediately gives an asymptotic period
$P_{\infty} \propto B_0^{(2-\alpha)/2}$ of the order of 170, 40, and 8 s, for
avenues A, B, and C respectively, assuming an initial field of 10$^{15}$ G.
In magnetars, the period $P_{\infty}$ is reached while the field strength is 
still larger than $\sim 10^{13}$ G, where equation (2) applies.

\subsection{Evolution in the $P - \dot P$ Diagram}%%%%%%%%%%%%%%%%%%%%%%%%%%%%%

\begin{figure}
\epsscale{1.0}
\plotone{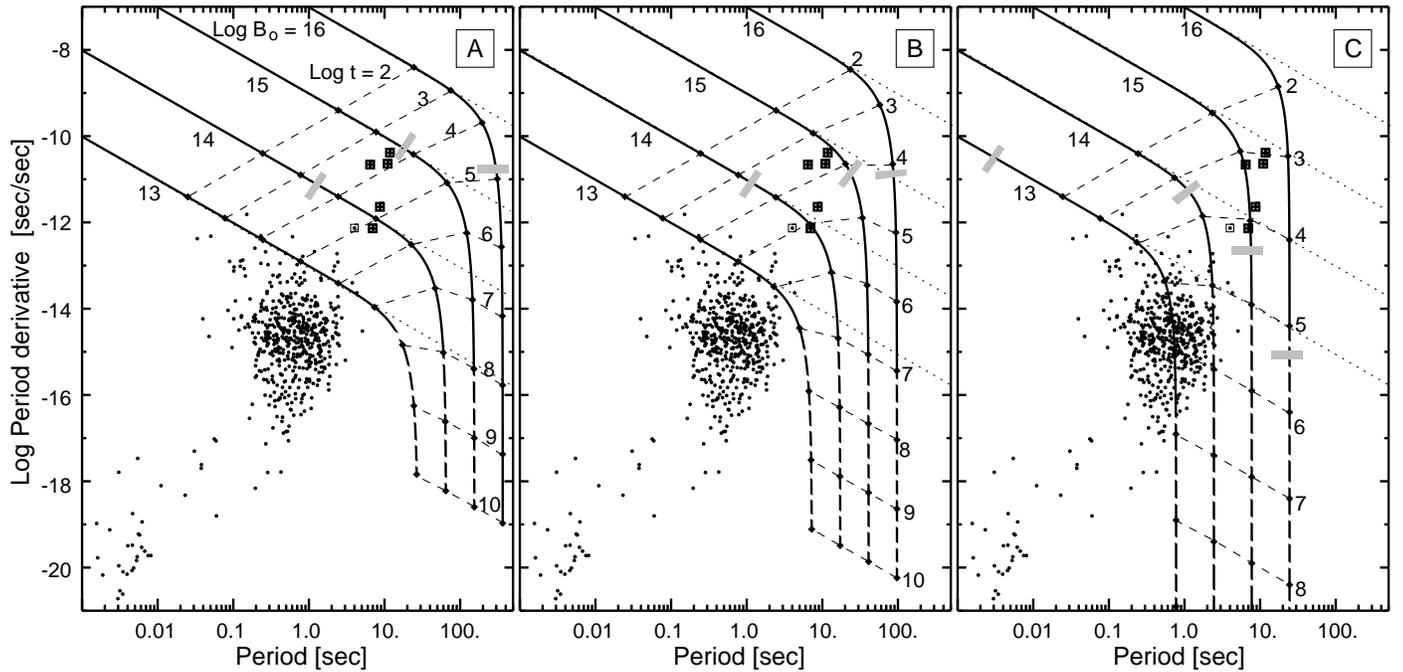}
\figcaption{Evolutionary tracks (solid lines) for avenues A, B, and C in the
$P-\dot{P}$ 
diagram.
The initial magnetic field strength $B_0$ is indicated for each curve and
the light dotted lines show evolution for a constant magnetic field.
On each track diamonds give the logarithm of the model age in years, as 
labeled,
and are connected by dashed lines for easier reading.
The grey bars indicate when our model thermal X-ray luminosity, estimated for 
neutron stars with iron envelopes, drops below 10$^{34}$ erg s$^{-1}$.
%***********DANY: DO DECIDE if you wish to say something like
%The evolutionary pathways are a good representation of the field
%decay as long as its strength remains above 10$^{13}$G.
%***********MONICA: I prefer to state it like this:
The evolutionary paths may not be reliable when the field drops below
10$^{13}$ G since our decay law (Eq.[\protect\ref{equ:decay-law}])
could be inapplicable below the magnetar regime: 
they are marked as broken in this region
since they could actually move to higher periods.
Radio pulsars from the Princeton catalogue 
(Taylor, Manchester, \& Lyne 1993\protect\markcite{TML93})
are plotted as dots and the boxed pulsar is
PSRJ1814-1744 (Camilo  et al. 1999\protect\markcite{Cetal99}). 
The five AXPs, shown as boxed crosses, in order of  decreasing period 
derivatives
are 1E 1841-045 (Vasisht \& Gotthelf 1997\protect\markcite{VG97}), 
1RXS J170849-40091 (Sugizaki et al. 1997\protect\markcite{Sugietal97}; 
Israel et al. 1999\protect\markcite{Isra99}),
1E 1048-5947 (Oosterbroek et al. 1998\protect\markcite{Oetal98}), 
4U 0142+61 (Israel et al. 1994\protect\markcite{Ietal94}), 
and finally 1E 2259+586 (Parmar et al. 1998\protect\markcite{Petal98}).
%***********DANY: IF WE DROP THEM we should remind ourself to cancel
% *MONICA: Yes, drop them.
%The two crosses denote 
%SGR 1900+14 (Woods et al. 1999\protect\markcite{Wetal99}) and 
%SGR 1806-20 (Kouveliotou  et al. 1998\protect\markcite{Ketal98}).
\label{fig:2}
}
\end{figure}

Figure~\ref{fig:2} illustrates tracks of the rotational evolution in the three 
field decay scenarios on $P - \dot P$ diagrams, comparing them with the 
observed
values for the five AXPs.

Notice that the time $t$ appearing in equation~(\ref{equ:period}) is the 
`real' age
of the star (which we will call the model age $t_{\rm mod}$).
While the field has not yet decayed and the period is much larger than the 
initial
period, the model age coincides with the spin-down age $t_{\rm sd} \equiv 
P/2\dot P$
but later on $t_{\rm mod} < t_{\rm sd}$, a general feature of any field decay 
model.

In case A all observed AXPs would have a magnetic field still very close to its
initial value and thus their ages are given by the spin-down age.
In case B the field decay proceeds faster than in A and the observed AXPs 
would be
approaching the stage of power-law field decay while in case C
(the fastest decay scenario) their fields would already have undergone some 
decay.
In the latter case the model ages are hence much lower than the spin-down
ages,
particularly for the lower field AXPs for which $t_{\rm mod} \sim 10^4$ yr 
while
$t_{\rm sd} \sim 10^5$ yr.

A second striking difference between the avenues A and B against C concerns the
location of the observed AXP periods relative to $P_\infty$.
While in A and B AXPs will still undergo a noticeable period increase, in C the
whole class of objects would have already attained the saturation period. 
Finally, it is interesting to notice that within avenue C all AXPs seem to be 
born
within a much narrower distribution of $B_0$ than in avenues A and B.

\subsection{Thermal X-Ray Luminosity}%%%%%%%%%%%%%%%%%%%%%%%%%%%%%%%%%%%%%%%%%%
%%%%%%

We now discuss the X-ray detectability of thermal emission from our model 
sources.
In cases A and B these X-ray luminosities have been calculated by 
H\&K\markcite{HK98}
and we here use their results, while for case C we use the results of Geppert 
et al.
(1999\markcite{Getal99}).

A basic feature of the thermal evolution with decaying magnetic field is that
longer decay time scales imply longer lifetimes for detectable thermal X-ray
luminosities which are then powered exclusively by the field decay.
Accordingly, avenue A may keep luminosities above 10$^{31}$ erg s$^{-1}$ 
(i.e., surface temperatures around $3 - 4 \cdot 10^5$ K) for a few
million years, while avenues B and C for just about one million years.
This luminosity would allow the model sources to be detected within a
few hundred parsecs.

In contradistinction, during the early phases, $\sim 10^3 - 10^4$ yr, 
when the thermal luminosity is above $10^{34}$ erg s$^{-1}$, the differences
between models A and B as compared to C are not so large.
This is due to a delicate balance between heating from field decay and 
neutrino losses
together with the initial heat content of the star and the transmission 
properties
of the strongly magnetized envelope.
We take the reference value of $10^{34}$ erg s$^{-1}$ as a lower limit for the 
visibility of the thermal component of AXPs. 
In each model the position where this luminosity is attained is 
marked in Figure 2 by a grey bar on the evolutionary tracks. 
All models assumed an iron envelope.
For light element enriched envelopes, the early thermal evolution would not be 
affected
but the observable thermal luminosities would rise by almost one order of 
magnitude.
The grey bars in Figure 2 would then correspond to about $10^{35}$ erg 
s$^{-1}$.

\section{DISCUSSION}%%%%%%%%%%%%%%%%%%%%%%%%%%%%%%%%%%%%%%%%%%%%%%%%

To address the question raised in the introduction, 
we now want to combine our above results on the rotational evolution and
on the thermal evolution.
We do not consider explicitly the SGRs in this work since their spin-down
is certainly not as steady as for AXPs and the use of the simple 
magnetic dipolar radiation braking law may be strongly misleading
(see, e.g., Harding, Contopoulos \& Kazanas 1999\markcite{HCK99}).
In contradistinction, 1E 1841-045 has shown an extremely regular spin-down
over more than ten years (Gotthelf, Vasisht \& Dotani 1999\markcite{GVD99})
and the ``bumpy'' spin-down observed in 1E1048.1-5937 and 1E 2259+586
can be interpreted within the magnetar hypothesis assuming a
``radiative precession'' superposed to the standard
magnetic dipole radiation braking
(Melatos 1999\markcite{mela99}; see also Kaspi, Chakrabarty \& Steinberger
1999\markcite{KCS99}).

The two sources 1E 2259+586 and 4U 0142+61 have model ages around 10$^5$ yr
in both avenues A and B, and hence their predicted thermal X-ray luminosities 
would be
much fainter than observed, as indicated by the location of the grey bars in
Figure 2.
On the other hand, in the case of avenue C these objects are much younger and 
thus
bright enough to be compatible with the observations.
One possibility to reconcile A and B with observations is to invoke the 
presence
of H/He envelopes in these two stars to raise their predicted 
thermal luminosity.
However, this may raise another problem when interpreting the evolutionary 
tracks
for the higher field of 10$^{15}$ G.
If we also assume the presence of light element envelopes in these paths, 
this would imply a shift in the limiting age for X-ray detectability from 
around 10$^4$
yr up to at least 10$^5$ yr (i.e., a shift of the grey bar in Figure 2) 
due to the action of the decaying field.
The first consequence of this shift would be a large enhancement of the 
detection
probability of sources on these tracks while only three AXPs are located there,
1E 1048.1-5937, 1E 1841-045 and 
1RXS J170849-40091, and all three are very young. 
Second, along these tracks the period would still be increasing toward the 
asymptotic value and would exceed the presently observed ones.
This problem is particularly severe for avenue A.
However, within avenue C the five AXPs are well within the predicted X-ray 
detectability range and there is no need to invoke the presence of a light
element envelope. 
Moreover, even if light elements were present they would not significantly 
raise
the limiting age for detectability because of the much faster 
cooling in avenue C, a result of the much faster field decay.

As a last remark we want to notice that scenario C hints for AXPs being a 
class of neutron stars born with initial magnetic field within a very
narrow strip around $10^{15}$ G. 
As a consequence, the young AXPs  1E 1841-045, 1E 1048-5947 
and 1RXS J170849-40091, with model age of 
10$^3$ yr, appear as the precursors of the 
older ones 4U 0142+61 and 1E 2259+586, 
with model age of 10$^4$ yr.
In contrast, for a not so fast decayed field, as in cases A and B, it is
difficult to explain the absence of the younger, and 
consequently X-ray brighter,
precurring sources of 4U 0142+61 and 1E 2259+586, the last two with 
model age $10^5$ yr.

\section{CONCLUSIONS}%%%%%%%%%%%%%%%%%%%%%%%%%%%%%%%%%%%%%%%%%%%%%%%%

In this letter we show that within the fastest field decay illustrated by
avenue C, all five AXPs are located on tracks which have already reached their
asymptotic period $\sim$ 8 seconds. This constitutes a natural explanation of 
the
observed period clustering of the five AXPs with measured period derivatives
as well as of the AXP candidate AX J1845-0258.
On the contrary, within the slower field decay scenarios A and B,
this clustering of periods appears more contrived.
We will now mention several consequences of this proposed scenario C that may
in future be tested.

RXJ 0720.4-3125, a recently discovered nearby isolated neutron star 
(Haberl et al. 1997\markcite{Hetal97}) with a rotation 
period of 8.39 s and purely thermal X-ray luminosity of 10$^{31}$ erg s$^{-1}$ 
is a
candidate for a middle aged magnetar 
(Heyl \& Hernquist 1998\markcite{HH980720}; H\&K\markcite{HK98}).
Heyl \& Hernquist considered constant magnetic field models while 
H\&K\markcite{HK98}
 used their results within avenues A and B, and interpreted the observed X-ray
luminosity as powered by the dissipation of the magnetic energy.
Within the interpretations of these authors, 
this source should have a field at birth slightly lower than the one of the 
AXPs
to explain that its period is only 8.39 s despite of its age, 
and a  period derivative (not yet determined) of a  few times 10$^{-13}$ s 
s$^{-1}$.
Avenue C nicely predicts the existence of a population of objects similar to
RXJ 0720.4-3125, with periods around 8 s or slightly longer, but with smaller
derivative.
At ages about 10$^5$ - 10$^6$ yr they have a luminosity of the order of
10$^{31}-10^{32}$ erg s$^{-1}$ and period derivatives between 10$^{-14}$ and
10$^{-16}$ s s$^{-1}$.
The measurement of $\dot P$ for RXJ 0720.4-3125 will enable us to discriminate 
among the different avenues.
Notice that if the source is rather powered by accretion from the interstellar 
medium
(Wang 1998\markcite{W98}; Konenkov \& Popov 1997\markcite{KP97}) 
then its period derivative should be below $10^{-16}$ s s$^{-1}$.

The two AXPs 1E 1841-045 and 1E 2259+586 are associated with the SNRs
Kes 73 and CTB 109 respectively 
(Gotthelf \& Vasisht  1997\markcite{GV97}; Parmar et al. 
1998\markcite{Petal98}).
The sources spin-down ages and the ages of the SNRs are in discrepancy
by factors of $\sim 2.3$ and $10$, respectively.
The field decay in avenue C is able to cure this problem naturally, but not A 
and B,
as is clear from Figure 2.

A related interesting result of avenue C concerns the recently discovered
radio pulsar PSR J1814-1744 (Camilo  et al. 1999\markcite{Cetal99}).
With a 4 s period and a measured $\dot {P}$ of $7.4 \cdot 10^{-13}$ s s$^{-1}$,
this radio pulsar is located, surprisingly, 
very close to the prototype AXP 1E 2259+586.
If we try to include it as a radio-loud AXP 
(Pivovaroff et al. 1999\markcite{PKC99}), avenue C would predict
an initial field somewhat below $10^{15}$ G and a model age of $\sim 10^4$ yr.
This opens the possibility to detect the associated SNR.
Within avenues A and B its age would be close to its spin-down age, i.e.,
$85000$ years.

Our conclusion is thus that the dipolar component of the magnetic field of 
highly magnetized
neutron stars must have a very short decay time scale.
The crustal magnetic field hypothesis on
which avenue C is based
gives the indication  that the  decay time scale, in magnetars, 
is actually controlled by physical processes occurring in the crust. 
Assuming that strong magnetic fields permeate the stellar core, we can only
conjecture that this may imply the action of 
some efficient mechanism of flux expulsion 
from the core into the crust, possibly in the line described, for example, by
Srinivasan et al (1990\markcite{Setal90}).

%%%%%%%%%%%%%%%%%%%%%%%%%%%%%%%%%%%%%%%%%%%%%%%%%%%%%%%%%%%%%%%%%%%%%%%%%%%%%%%
%%%
%%%%%%%%%%%%%%%%%%%%%%%%%%%%%%%%%%%%%%%%%%%%%%%%%%%%%%%%%%%%%%%%%%%%%%%%%%%%%%%
%%%
%%%%%%%%%%%%%%%%%%%%%%%%%%%%%%%%%%%%%%%%%%%%%%%%%%%%%%%%%%%%%%%%%%%%%%%%%%%%%%%
%%%
%%%%%%%%%%%%%%%%%%%%%%%%%%%%%%%%%%%%%%%%%%%%%%%%%%%%%%%%%%%%%%%%%%%%%%%%%%%%%%%
%%%

%%%%%%%%%%%%%%%%%%%%%%%%%%%%%%%%%%%%%%%%%%%%%%%%%%%%%%%%%%%%%%%%%%%%%%%%%%%%%%%
%%%
%%%%%%%%%%%%%%%%%%%%%%%%%%%%%%%%%%%%%%%%%%%%%%%%%%%%%%%%%%%%%%%%%%%%%%%%%%%%%%%
%%%
%%%%%%%%%%%%%%%%%%%%%%%%%%%%%%%%%%%%%%%%%%%%%%%%%%%%%%%%%%%%%%%%%%%%%%%%%%%%%%%
%%%
%%%%%%%%%%%%%%%%%%%%%%%%%%%%%%%%%%%%%%%%%%%%%%%%%%%%%%%%%%%%%%%%%%%%%%%%%%%%%%%
%%%

\bigskip
{\bf Acknowledgments.} The authors would like to thank A. Possenti and
T. Zannias for stimulating discussions, and to E. Gotthelf and S.
Mereghetti for a critical reading of the manuscript.  M.C. and D.P. are
thankful to
the Astrophysikalisches Institut Potsdam for kind hospitality  during the 
preparation of
this work.
This work was supported by Conacyt (\#27987E), UNAM-DGAPA (\#IN119998) 
and a binational grant DFG (\#444 - MEX - 1131410) - Co\-na\-cyt.

\bigskip
\newpage
%%
%%%%%%%%%%%%%%%%%%%%%%%%%%%%%%%%%%%%%%%%%%%%%%%%%%%%%%%%%%%%%%%
%%
%%%%%%%%%%%%%%%%%%%%%%%%%%%%%%%%%%%%%%%%%%%%%%%%%%%%%%%%%%%%%%%%%%%%%%%%%%%%%%%
%%

%%%%%%%%%%%%%%%%%%%%%%%%%%%%%%%%%%%%%%%%%%%%%%%%%%%%%%%%%%%%%%%%%%%%%%%%%%%%%%%
%%
\newpage
% FIGURES
%%%%%%%%%%%%%%%%%%%%%%%%%%%%%%%%%%%%%%%%%%%%%%%%%%%%%%%%%%%%%%%%%%%%%%%%%%%%%%%
%%

%%%%%%%%%%%%%%%%%%%%%%%%%%%%%%%%%%%%%%%%%%%%%%%%%%%%%%%%%%%%%%%%%%%%%%%%%%%%%%%
%%
%%%%%%%%%%%%%%%%%%%%%%%%%%%%%%%%%%%%%%%%%%%%%%%%%%%%%%%%%%%%%%%%%%%%%%%%%%%%%%%
%%

\begin{references}%%%%%%%%%%%%%%%%%%%%%%%%%%%%%%%%%%%%%%%%%%%%%%%%%%%%%%%%%%%%%
%%
%%%%%%%%%%%%%%%%%%%%%%%%%%%%%%%%%%%%%%%%%%%%%%%%%%%%%%%%%%%%%%%%%%%%%%%%%%%%%%%
%%
%%%%%%%%%%%%%%%%%%%%%%%%%%%%%%%%%%%%%%%%%%%%%%%%%%%%%%%%%%%%%%%%%%%%%%%%%%%%%%%
%%

\reference{Cetal99}
Camilo, F., et al. 1999, Nature, submitted

\reference{Coe98}
Coe, M.J., \& Pightling, S.L. 1998, MNRAS, 299, 223

\reference{DT92}
Duncan, R. C., \& Thompson, C. 1992, ApJ, 392, L9

\reference{Gaen98}
Gaensler, B.M., Gotthelf, E.V., \& Vasisht, G. 1999, 
ApJ, 526, L37

\reference{Getal99}
Geppert, U., Page, D., Colpi, M., \& Zannias, T. 1999, to appear in 
the proceedings of the IAU Colloquium 177
(e-print: astro-ph/9910563)

\reference{GV97}
Gotthelf, E.V., \& Vasisht, G. 1997, ApJ, 486, L133

\reference{GV98}
Gotthelf, E.V., \& Vasisht, G. 1998, New Astronomy, 3, 293

\reference{GVD99}
Gotthelf, E.V., Vasisht, G., \& Dotani, T. 1999,
ApJ, 522, L49

\reference{GR92}
Goldreich, P., \& Reisenegger, A. 1992, ApJ, 395, 250

\reference{Hetal97}
Haberl, F., et al. 1997, A\&A 326, 662

\reference{HCK99}
Harding, A. K., Contopoulos, I., \& Kazanas, D. 1999,
ApJ, 525, L125

\reference{HH980720}
Heyl, J. S. \& Hernquist, L. 1998, MNRAS, 297, L69

\reference{HH98ultra}
Heyl, J. S., \& Hernquist, L. 1998, MNRAS, 300, 599

\reference{HK98}
Heyl, J. S., \& Kulkarni, S. R. 1998, ApJ, 506, L61

\reference{Ietal94}
Israel, G.L., Mereghetti, S., \& Stella, L. 1994, ApJ, 433, L25 

\reference{Isra99}
Israel, G.L., et al. 1999, ApJ, 518, L107

\reference{KCS99}
Kaspi, V.M., Chakrabarty, D., \& Steinberger, J. 1999, ApJ, 525, L33


\reference{KP97}
Konenkov, D. Yu., \& Popov, S.B. 1997, AstL, 23, 498


\reference{mela99}
Melatos, A. 1999,
ApJ, 519, L77

\reference{MS95}
Mereghetti, S., \& Stella, L. 1995, ApJ, 442, L17 

\reference{MIS98a}
Mereghetti, S., Israel, G. L., \& Stella, L. 1998, MNRAS, 296, 689

\reference{Oetal98}
Oosterbroek, T.,  et al. 1998, A\&A, 334, 925

\reference{P98}
Page, D. 1998,
in Neutron Stars and Pulsars,
eds N. Shibazaki, N. Kawai, S. Shibata, \& T. Kifune, 
(Universal Academy Press: Tokyo), p. 183.

\reference{Petal98}
Parmar, A. N., et al. 1998, A\&A, 330, 175

\reference{PKC99}
Pivovaroff, M., Kaspi, V. \& Camilo F. 1999, to appear in 
the proceedings of the IAU Colloquium 177

\reference{Setal90}
Srinivasan, G., Bhattacharya, D., Muslimov, A. G., \& Tsygan, A. I. 1990,
Curr. Sci., 59, 31

\reference{Sugietal97}
Sugizaki, M., et al. 1997, PASJ, 49, L25

\reference{TML93}
Taylor, J. H., Manchester, R. N., \& Lyne, A. G. 1993,
ApJS, 88, 529

\reference{TD96}
Thompson, C., \& Duncan, R. C. 1996, ApJ, 473, 322

\reference{TDetal98}
Torii, K., et al. 1998, ApJ, 503, 843

%\reference{U97}
%Usov, V. V. 1997, A\&A, 317, L87

\reference{VG97}
Vasisht, G., \& Gotthelf, E. V. 1997, ApJ, 486, L129

\reference{W98}
Wang, J.C.L. 1998, ApJL, 497, L55

\reference{W64}
Woltjer, L. 1964, ApJ, 140, 1309


%%%%%%%%%%%%%%%%%%%%%%%%%%%%%%%%%%%%%%%%%%%%%%%%%%%%%%%%%%%%%%%%%%%%%%%%%%%%%%%
%%
\end{references}
\end {document}